\documentstyle[aps,prl,multicol,epsf]{revtex}

\begin{document}

\title{Switching current noise and relaxation of ferroelectric domains }
\author{Bosiljka Tadi\'c }

\address{Jo\v{z}ef Stefan Institute,
P.O. Box 3000, 1001 Ljubljana, Slovenia  }


\maketitle
\begin{abstract}
\newline
We simulate field-induced  nucleation and switching of domains in a
 three-dimensional  model of ferroelectrics with quenched  disorder
and varying domain sizes. We study (1) bursts of the switching current
at  slow driving  along the  hysteresis loop  (electrical Barkhausen noise)
 and (2) the  polarization reversal when a strong electric field was
 applied and  back-switching after the field was removed.
We show how these processes  are related to the  underlying structure of
domain walls, which in turn is  controlled by the  pinning at quenched
local electric fields.
 When the depolarization fields of bound charges are properly screened
we find that the fractal switching current noise may appear with two
distinct universal behaviors. The critical depinning of plane domain walls
determines the universality class in the case of  weak random fields,
whereas for large randomness the  massive nucleation of
domains in the bulk leads to different scaling properties.
 In both cases the scaling exponents decay logarithmically  when the
driving frequency is increased.
The polarization reverses in the applied  field as a power-law, while
its relaxation in zero field  is a stretch exponential function of time.
The stretching  exponent depends  on the strength of pinning. The results
may be applicable for uniaxial relaxor ferroelectrics, such as doped SBN:Ce.

\end{abstract}

\begin{multicols}{2}

\newpage

\section{Introduction}

Polarization reversal and  properties of the accompanying  switching
current  are relevant  for various applications of ferroelectric materials.
In particular, it is a key process that takes place in electronic devices
based on ferroelectrics thin films, where a reproducible switching process is
required.
Apart from the extensive experimental and theoretical work
\cite{Rudyak,book,Shur_rev}, a complete understanding of the domain-wall
motion and predictability of the switching processes remains elusive in
real ferroelectrics, especially  when several physical parameters,
determining either structure or dynamics, are varied.
Domain switching and relaxation (back-switching) in ferroelectrics can be
studied by hysteresis loop, switching current, and switching current
bursts (electrical Barkhausen noise).

The development of domain structure and motion of domain walls  during
switching can be regarded in the scope of nonlinear dynamic  processes
occurring in driven disordered systems, which are studied extensively in
the past decade.
Particularly, the measurements on ferromagnets (a short review is given in
\cite{BT_summary}) show that Barkhausen noise exhibits interesting fractal
properties, that can be characterized by a full set of scaling exponents
of the sizes, durations and energies of the bursts and of the power spectrum.
An example with a  complete set of measured scaling exponents and
corresponding fractal dimensions with  derivation of the associated
scaling relations can be found in Ref.\ \cite{Djole}.
A possible grouping of the scaling exponents into different
universality classes of scaling behavior  offers a new and
simple link between the  measured Barkhausen  noise
and the domain structure. For example, in  a recent study  \cite{packing}
such a  relation was established  between the universal scaling exponents of
acoustic emission noise and the packing properties near the first-order
structural transformation.

 The electrical Barkhausen noise  has been observed experimentally in
${\mathrm{BaTiO_3}}$ \cite{Rudyak,book,BTiO3} and
${\mathrm{Gd_2(MoO_4)_3}}$ and ${\mathrm{LiTaO_3}}$
\cite{Shur,Shur_et_al}  single crystals and
in  ${\mathrm{9.4/65/35 (Pb,La)(Zr,Ti)O_3}}$ (PLZT) ceramics
\cite{ceramics} and  ${\mathrm{PbMg_{1/3}Nb_{2/3}O_3}}$ (PMN)
relaxor ferroelectric \cite{Weissmann}.
Physically, the polarization reversal and domain wall motions in
ferroelectrics is accompanied with the local change of sign of bound  charges,
which influences values of the electric field in the sample. The
depolarization field of bound charges can be partially screened by an
adequate addition of charges on the surface and by
other methods (see \cite{Shur_rev} and references therein).
Hence at a given point $x$ in the sample at time $t$ un-compensated
depolarization field is given by
\cite{Shur_rev}
\begin{equation}
\delta E_{dep}(x,t)=
E_{dep}(x,t)-\sum E_{sreen}(x,t)\ ,
\label{eff_field}
\end{equation}
where $\sum E_{sreen}(x,t)$ is a total field due to screening (surface and
bulk)  charges at that site. Incomplete screening of the depolarization field
tends to change  the electric field inside the sample, thus
influencing the course of switching process itself. For instance, there is
a finite probability that inverse jumps of polarization (negative
Barkhausen pulses) \cite{Rudyak,Shur_rev} occur, that slow down the switching.
Recently a series of sophisticated measurements  were done
\cite{Shur_et_al,FTT} that reveal the effects of the depolarization field on
a  created plane domain wall in ${\mathrm{Gd_2(MoO_4)_3}}$ and on a
multi-domain structure in ${\mathrm{LiTaO_3}}$.
The obtained  distributions of lengths, sizes and rest time of Barkhausen
jumps obey a  power-law behavior with an exponential cut off, both for
switching and back-switching processes.
Using a suitable experimental set-up  the power-spectrum in PMN relaxors
was  measured, which was
found to  decay with frequency $\omega $ as  $S(\omega) \sim \omega^{-2}$
\cite{Weissmann}.

Recently studied uniaxial relaxor ferroelectrics
${\mathrm{Sr_{0.61}Ba_{0.39}Nb_2O_6}}$  (SBN61) and
${\mathrm{Pb_{0.61}Ba_{0.39}Nb_2O_6}}$ are regarded as
systems with the Ising symmetry of the order parameter,  having the tetragonal
structure  in the low-temperature phase with the spontaneous polarization
along c-axis  \cite{Dui2,Peer}.
 In solid solutions ${\mathrm{SBN61:X}}$ doping with ${\mathrm{X=Ce^{3+},
Cr^{3+},  Co^{2+}}}$ reduces the transition temperature and induces
 local electric random fields, which give rise to local polar clusters
\cite{Peer}.
Recently a comprehensive series of measurements in SBN:Ce and similar
systems \cite{Dui2,Peer,Kleemann}  gave a strong support  to
the presence of random electric fields in this class of relaxor
ferroelectrics.
In general, more common sources of uncorrelated random fields in
ferroelectrics are neutral defects that can locally break the polar symmetry,
as for instance in TGS (see detailed discussion in Ref.\ \cite{book_AL},
page 149).
[Note that, in contrast to ferroelectrics,  in ferromagnets local
random fields are not
allowed by the symmetry, but are theoretically constructed as being induced
by coarse graining due to other types of disorder (see also \cite{BT_letter}).]

In this work we simulate  temporal behavior of polarization driven by the
electric field  in three-dimensional systems of Ising spins subject to
{\it quenched} local random  fields and  positive nearst-neighbor
interaction, which is
compatible with the occurrence of a ferroelectric long-range order.
 We consider two types of driving.  First, by stepwise increase
(decrease) of the external field along a
hysteresis loop (see below) we study  bursts in the switching current
$j(t) = dP(t)/dt$
(electrical Barkhausen noise). We discuss effects of incomplete screening
of depolarization fields on the coarse of switching and hysteresis loop
properties. When the screening is complete, we study  how the
statistical properties of the noise depend on the strength of
pinning and on the frequency  with which the driving field is  changed.
Second,  by imposing a high opposite value of the external field in the
homogeneous state we let the polarization to reach its saturation at
that field, and then remove the field. We determine
reversal and spontaneous back-switching of the polarization with time at
different strength of disorder.
This type of switching is  more familiar in the experiments
on ferroelectrics \cite{Kleemann}.

Apart from allowing negative Barkhausen pulses as described below,
the present study is complementary compared to earlier simulations of
Barkhausen pulses in 3-dimensional Ising systems in the presence of disorder
\cite{BT_ferroel,BCN,Cornel,Zheng} in the following sense: In the earlier
work  \cite{BT_ferroel} we considered only strong disorder that is
compatible with
nucleation of small domains under driving  field, as for instance in
relaxor ceramics, and  analyzed the signal from only central part
of the hysteresis loop, as it is usually done on experiments.
In the present work  the study is extended to entire range of disorder
including the weak pinning region where extended domain walls may occur
due to lack of pinning centers \cite{comment_planarDW}.
In this case a potential depinning of domain walls at a critical driving
field contributes significantly to the scaling behavior of Barkhausen
pulses, as recent theoretical results on Bethe lattice suggest \cite{DD}.
Therefore, we include integration over entire hysteresis loop
in order to incorporate the contribution of critical depinning to the
universal scaling properties of the  induced electrical
noise in the case of  weak pinning \cite{comment_integration}.
Furthermore, here we employ the {\it simulation algorithm at finite driving
rates}, that mimics  closely the experimental situation with stepwise
increase (decrease) of the electric field.
We examine in detail the effects of driving rates in both strong and
weak pinning case by varying height of the steps in the external field.
The earlier simulations  in Refs.\ \cite{BCN,Cornel} were done in the
limit of zero driving rate, which appears physically inaccessible to
real experiments.  Whereas an entirely different
aspect concentrating on the  dynamic scaling during the growth of {\it
individual} avalanches was  considered in \cite{Zheng}.
It should be stressed that in this paper we implement strictly {\it quenched
disorder}, as described in Section II. Note that some speed algorithms,
as for instance used in  \cite{Cornel,Zheng}, allow certain  dynamic
variations of pinning  while the avalanche grows, which  can be  relevant
to  systems with annealed  rather than quenched disorder.
Theoretically, having unfixed pinning centers during the avalanche growth
may alter the long-range correlations in the avalanche-triggering fields
and in distances between initial points of avalanches,  which were shown
to be closely associated with the appearance of the fractal noise in
two-dimensional systems  \cite{correlations}.

The paper is organized as follows. In Sec.\ II we introduce the model and
discuss all relevant parameters that influence the hysteresis loop properties.
In Sec.\ III we study in detail scaling behavior of the switching current
noise under varying disorder and driving conditions.  Sec.\ IV contains the
results  of switching in a high field and
spontaneous back switching  of polarization when the pinning is varied.
The paper contains  a short summary of the results and discussion in Sec.\ V.

\section{Model and relevant parameters }

We consider a model with Ising-type  spins situated on three-dimensional
simple cubic lattice with positive interactions between neighbor spins and
local random field at each lattice site \cite{BT_letter}:
\begin{equation}
{\cal{H}} = -\sum _{x,y}J_{xy}S(x)S(y) - \sum _xh_xS(x)  -E\sum _xS(x)\ .
\label{Hamiltonian}
\end{equation}
We assume that the random exchange interactions $J_{x,y}$ are
distributed around a
positive value $J$ with a narrow distribution. The random fields $h_x$
are given by Gaussian distribution  with  zero mean and the width $f$
(measured in units of $J^2$). As mentioned above, the source of local
random fields can be  neutral defects that break polar
symmetry \cite{book_AL} or charge disorder due to
doping \cite{Peer},  which cause occurrence of polar regions in
relaxor ferroelectrics.
In addition, a random part of the interaction between these
polar regions  contributes to the observed glassy behavior of the
dielectric response \cite{Rasa}. Since the random
fields locally break the symmetry of the order parameter, we believe
that the dominant disorder effects on the
dynamics that we consider are due to random fields. Therefore,  here we
neglect the randomness in the spin-spin interaction.

The spin dynamics that we are considering is field-assisted and takes part
deeply in the ordered phase. Hence, we neglect possible temperature
effects, both for the reasons of keeping the number of parameters finite
and assuming that they play a secondary role in these processes.
The dynamics consists of spin alignment along a locally dominant field.
Thus, when the local field at site $x$ and time $t$, $h_{loc}(x,t)$ exceeds
zero, the spin $S(x,t)$ at that site flips and thus aligns along the local
field. The local field  $h_{loc}(x,t)$ consists of the interaction and pinning
part  $h_{ip}(x,t)= \sum_{y}J_{xy}S(y,t) +h(x)$  and
the driving field  $E(x,t)$. The driving field itself is  given by
$E(x,t) = E_{ext}(t) - \delta E_{dep}(x,t) $, where  $E_{ext}(t)$ is
the external field  and $\delta E_{dep}(x,t) $ a non-compensated
part of the depolarization field at that site \cite{Shur_rev}.
We employ a slow stepwise increase of the  external field
$E_{ext}(t_i)=-E_{sat}+ t_i\Delta E$ at discrete time intervals $\{t_i\}$,
while assuming that a non-compensated
part of the depolarization field at site $x$ may assist a probabilistic
(probability parameter $b$)  back-flip of an already aligned spin at that
site.
It should be stressed that the parameter $b$ represents the probability
that at a site $x$ in time $t$ the depolarization field opposes the effect
of external filed.
(The situation where the  depolarization field is parallel to the external
field is less interesting within the present numeric implementation of
the dynamics where driving rate is not fixed, as described below.)
Having in mind that the depolarization field at a given site depends on
the actual domain structure and its kinetics, the uniform distribution of
the stochastic variable $b$ can be regarded only as a first step towards
modeling of this complex phenomena.

Implementation of the dynamics is as follows:
Quenched random fields of Gaussian distribution (in double precision)
are generated and stored. They are {\it kept constant until entire
 hysteresis loop is completed} by the driving field. The external field is
increased by a small amount $\Delta E$ and the set of local fields
$\{h^{loc}(x,t)\}$ computed at each lattice site  and stored.
The system is updated {\it in parallel}, i.e., with respect to stored
set of local fields, and the spin flipped when
$h^{ip}(x,t) + E_{ext}(x,t)$ exceeds zero by a small amount $ 10^{-10}$.
When it is assumed that the depolarization  field is not entirely compensated,
the spin flip is executed with a reduced probability $1-b$.
This completes one time step of avalanche evolution.
Then a new set of local fields at affected sites (neighbors of just flipped
spins) is computed and stored. The conditions for spin flips at those sites
examined and spins flipped according to the above rule.
The process continues  until no more spins are found which satisfy the
conditions for flip. This completes one
Barkhausen avalanche. Then the external field is increased again by
the amount $\Delta E$.
When the  whole loop is completed, we set another distribution of quenched
random fields and repeat the driving. The results are averaged over the entire
ensemble (up to 200) of different random-field realizations.

In this model the following relevant parameters appear: Width of the
random-field distribution $f$, that determines strength of pinning and/or
domain size; jump $\Delta E$ of the external field, which  defines the
driving rate; probability of back-flips $b$,  which is related to the
 unscreened  part of the depolarization  field. In practice,
 usually an additional parameter is due to the section of
the hysteresis loop at which the signal is monitored, since the properties of
the signal vary along the loop. As mentioned above, here we integrate the
statistics over the entire hysteresis loop in order to capture the
contribution of the depinning of large domain walls to the scaling behavior
of noise by  weak pinning.

The hysteresis loop dependence on pinning by random fields and
on driving rates  is shown in Fig.\ 1. By varying the strength of
pinning $f$  the form of the hysteresis loop changes (Fig.\ 1a) from the
rectangular loop at low disordered 3-d sample  to the
slim loop, when the pinning is strong (large $f$). In the case of low
disorder, i.e., at weak pinning of domain walls, large domains may form
due to lack of pinning centers. In real ferroelectrics occurrence of
large domains are usually accompanied by extended plane domain walls
\cite{BTiO3,Shur,comment_planarDW}.
 On the other side, when the pinning is strong,
such as in the case of pinning at grain boundaries of small grain
ceramics \cite{ceramics}, many small domains can be nucleated at spatially
distributed centers with weak pinning, and their walls can propagate
along few active sites with weakest pinning, as the driving field increases.
This makes  effectively low-dimensional   walls moving in a
three-dimensional sample. In addition, a massive
nucleation of domains in the central part of hysteresis imposes  a spatial
restriction on the growth of later domains. Hence we expect a different
scaling behavior of sizes and durations of induced switched domains in
the case of weak and strong disorder. Our simulations confirm this
phenomenological picture.
In ferroelectrics a continuous variations  of the size of domains can be
achieved by increasing temperature or by applying  stress, as for
instance described
in  8/65/35 PLZT ceramics \cite{PLZT-HL}. It was suggested in Ref.\
\cite{Kleemann} that the appropriate illumination of samples can
notably reduce the strength of pinning in doped relaxor
${\mathrm{SBN61:Ce^{3+}}}$ ferroelectrics.

We would like to stress that in the algorithm  keeping the external
field constant during the
avalanche evolution and increasing it only  {\it after} the avalanche
has stopped does not allow to study rest periods between avalanches.
However, this type of field update leads to a better resolution of the noise,
while it changes the time dependence of the driving field (cf. Fig.\ 1c).
The driving rate can be better defined as the average derivative
$r\equiv <dE(t)/dt>$ along the loop.
In the simulations we find that  when back flips occur with a finite
probability  $b>0$ they effectively reduce increase of the driving field
and prolong the duration of a jump, as well as inducing inverse jumps,
as shown in  Fig.\ 1 c,d.

Assuming that the back flips occur very rarely ($b\to 0$),
we obtain the switching current noise which is shown in Fig.\ 2.
The train of bursts of  the switching current exhibits a long-range
correlations---Fourier spectrum $j(\omega )$ of the
signal is shown in Fig.\ 2 (top).  It is related to  the power-spectrum
via $S(\omega )=|j(\omega )|^2$, that can be directly measured
(see \cite{Weissmann}).
Hence, the exponent of the power spectrum in the present model is
given by $2\times \phi =1.48\pm 0.02$ for the case of moderate pinning
$f=3.6$. The power-law behavior
of the power spectrum suggests that the switching current noise exhibits
fractal character when  certain conditions are met, in particular, when the
depolarization fields are successfully screened.

\section{Fractal properties of the switching-current noise }

The fractal noise of the switching current have a range of universal
scaling properties, that can be determined more clearly if the field
inside the sample is fully controlled. Therefore, in this section we
put $b=0$ in order to study in detail these scaling
 properties and to determine how they depend on two relevant
parameters---the domain size and the driving rate.

\subsection{Varying strength of pinning}

First we fix the driving rate to a small value $\Delta E/J=1\times10^{-3}$
and consider how the domain size or strength of pinning influence the
properties of the switching current.
In Fig.\ 3 we show the cumulative  distributions of size (area) $s$ of
switched domains  and duration $t$ of switching following a single
field update, for several values of  pinning strength $f$.
As the Fig.\ 3 shows, both switched areas and durations of their switching
show a wide range of values that are described by  power-law
distributions with finite cut-offs ($X$ stands for  $X\equiv s,t$):
\begin{equation}
P(X,L) = X^{-(\tau_X-1)}{\cal{P}}(XL^{-D_X}) \ .
\label{DX}
\end{equation}
The scaling exponents $\tau _s-1$ and $\tau _t-1$ determine
the respective slopes of the cumulative distributions of  switched area
 and duration of switching (see Fig.\ 3).
 The fractal dimension $D_s$ of size of switched domains
and the  dynamic exponent  $D_t\equiv z$ in Eq.\ (\ref{DX})
determine how the  corresponding cut-offs scale with the characteristic
length (here the system size $L$),  $X_0 \sim L^{D_X}$.

Both the slopes $\tau_X$ and the cut-offs, related to $D_X$, of the
distributions depend on the strength of pinning.
First, for weak pinning (small values of $f$, bottom curve in Fig.\ 3) we
notice that extended domains, bounded by  presumably plane  domain walls,
 may occur.
By increasing the field these  domain walls move through a weakly
random medium and eventually
depin at a critical field $H_d(f)$. Depinning of large domain walls leads to a
peak in the histogram at large values of $s$, or a  plateau in the
cumulative distribution as shown
 in Fig.\ 3. Before the depinning is reached a series of small avalanches
occurs that are power-law distributed.
The plateau disappears at a critical value of pinning strength
$f_c(r,L)\approx 2.4$,
where a pure power-law distribution with virtually infinite cut-off (controlled
only by the system size) occurs. Here the measured slopes give $\tau_s=1.74$
and $\tau_t=2.14$ for the fixed driving rate $\Delta E/J=1\times 10^{-3}$
(see  Fig.\ 3).

At strong  pinning (large $f$) many small domains are being nucleated at
isolated points in the bulk and grow slowly along a few active sites per time
step.  The maximum sizes of domains are finite
and decrease with stronger pinning  (cf. top two curves in
Fig.\ 3). Here the measured slopes
give $\tau_s=1.56$ and $\tau_t=1.94$ for size and duration,
respectively, suggesting another universality class  of the scaling
behavior, compared to the case of weak pinning.
The transition to the strong pinning regime for $f > f_c$ is marked
by disappearance of the critical depinning of large domain walls at all
fields and domination of the nucleation processes.
Numerical data in Fig.\ 3 suggest that the critical value  $f_c$ is close to
 2.4 for the used  system size and driving rate.

Qualitatively similar behavior with two universality classes but with
different exponents, was found in the case of two-dimensional system
where an extended one-dimensional domain wall was initially prepared
\cite{TN}. For comparison, in Table\ 1 we summarize exponents both for
3-dimensional and for 2-dimensional case. (Notice that the listed
 exponents for three-dimensional system are obtained for a small finite
driving rate $\Delta E/J=4\times  10^{-4}$, whereas the simulations at zero
driving rate were  employed in two dimensions.)
The results for  2-dimensional case are expected to be
relevant for domains on the surfaces of the bulk samples (see Ref.
\cite{Peer}) and for  thin films in which the width of the domain wall
exceed the film thickness. An interesting physical realization  of
varying strength of pinning was suggested recently \cite{Co_film} in
ferromagnetic ${\mathrm C_o/C_oO}$ film by varying temperature through the
transition temperature at which the substrate orders
{\it anti-}ferromagnetically.
As already mentioned,  in ferroelectrics the local random fields are physical
\cite{Dui2,Peer,Kleemann,book_AL}. Therefore varying the strength of
pinning due to random fields in
ferroelectric thin films and bulk materials  can be achieved in a more
direct way, offering  diverse possibilities to detect the change in the
scaling behavior of the switching current noise and  critical properties
on the hysteresis.

\subsection{Varying driving rate}

 In the following we examine how the scaling properties of the switching
current noise change when the driving  rate is increased.
 Very low frequencies of the driving field are usually not accessible on
experiments, both for physical and technical reasons. A finite driving
rate, which is defined as $r\equiv <dE(t)/dt>$, can be approximated with
$\Delta E/J$ in the case of stepwise increase of the external field in the
absence of back flips $b=0$. For finite rate $r>0$ the switching may start
at different parts of the sample and the measured signal is a superposition
of these separate events. In contrast to the mathematical limit of
infinitely slow driving, by  finite field jump $\Delta E$  the domain wall
 may get enough energy to overcome several  pinning centers at the same time.
Thus either different small domains are initiated simultaneously, or
different parts of a large domain wall advance
in parallel. In both cases the statistical properties of the signal are
changed. Merging  of many  individual jumps makes the occurrence of large
events  more probable.

For a range of values of driving rates we still find a power-law behavior
with reduced values of the  scaling exponents and enlarged cut-offs.
In Fig.\ 4 we show how the distributions for sizes and durations of pulses
vary when the driving rate $\Delta E/J$ is increased, while the strength of
pinning was kept constant in the strong pinning regime. In Fig.\ 5 the
same type of analysis was done for the case of weak  pinning.
In the strong pinning regime a dramatic increase of the cut-offs
in the distribution of switched area occurs due to merging of many small
signals. Whereas, in the weak pinning the relative enlargement of the cut-off
is smaller, that can be understood as an effect of parallel advancement of
different sections of the same extended wall, rather than merging of
different domains.
Both in strong and weak pinning case the increased incidence of large events
leads to decrease of the scaling exponents. We find {\it logarithmic}
dependence
of the exponents for the size  and duration of
events on  $\Delta E/J$ both in weak and strong pinning regimes.
The results are shown in Fig.\ 6 and the fit lines are given by the
general form
\begin{equation}
\tau_X = A_X - B_X\ln (\Delta E/J) \ .
\label{logE}
\end{equation}
For the strong pinning both area and duration exponents decrease
equally, i.e.,  $B_s\approx B_t \approx 0.08$, with the constants
$A_s\approx 1.0$ and $A_t\approx 1.4$.
In the weak pinning, however, the duration exponent decreases with nearly
double rate compared to the size exponent. We find $B_t \approx 2B_s
\approx 0.11$, (see Fig.\ 6) and $A_t\approx 1.37$, $A_s \approx 1.30$
within numerical error bars.

It is important to note that the measuring conditions depend on the
experimental set up and they can vary from one experiment to the other.
The effective driving rate depends on the steps in the external field and
the degree of compensation of depolarization fields inside the sample.
In real experiments  the driving rate is always finite, resulting
in the scaling exponents that are always {\it smaller} (and the fractal
dimensions larger) compared to those computed in
the  theoretical limit of zero driving rate $r \to 0$.
Therefore, in order to compare measured and predicted scaling exponents
it is advisable to make several  measurements at  different driving rates and
find out if the results, for instance for the area exponent $\tau_s$, belong
to a line in the range 1.66---1.40, which is compatible with occurrence of
small domains, or in the range 1.78---1.60, suggesting that the sample has
extended plane domain walls.
When the driving rate is too large (cf. top line on both panels in
Fig.\ 5), the scaling behavior of the distributions is lost. The cumulative
probability distribution shows curvature both due to the lack of small
events and the excess of large events.

\section{ Polarization reversal and back-switching}

In the previous section we have shown how different kinds of domain walls
in the sample determine the scaling properties of  switching
current  noise, reflecting the nature of domain-wall motion when the electric
field is slowly ramped. Here we consider another type of driving.  First,
in the initially polarized   state (all spins down) we impose a large opposite
field $E_s$ and keep it until polarization reverses and reaches saturation
in that field.  Then we switch off the field to zero and monitor  how the
 polarization decays.  Both processes depend on the  type of domains in the
sample, and thus, on the strength of pinning.
The results for polarization as a function of time are shown in Fig.\ 7 for
various strengths of pinning $f$.

For weak pinning  spontaneous back switching of the polarization is
exponentially fast, reaching the value $P_r$ of remanent polarization.
Whereas, the relaxation becomes increasingly slow  as the strength of
pinning $f$ increases. The limiting  values of $P_r$ are lower, which
correspond to increasingly slimmer loops for strong
pinning  (cf. form of the hysteresis loop in Fig.\ 1a).  In general,
polarization decay in Fig.\ 7 can be approximated with a stretch
exponential law $P(t) = A+B\exp[-(x/\tau )^\sigma]$,
with $\sigma =\sigma (f)$  decreasing with the pinning strength. We have
$\sigma \approx $1, 0.92 and 0.80 for $f=$ 2.4, 3.4, and 4.4, respectively.
Qualitatively similar back-switching
was recently observed in doped  ${\mathrm{SBN:Ce^{3+}}}$ relaxor ferroelectric
 with two different stretching exponents corresponding respectively  to
illuminated and non-illuminated sample \cite{Kleemann}.

The polarization reversal in the strong constant field shows slow increase
in the presence of small domains, but with another type of time dependence
(see Fig.\ 7).  The positive part of the polarization
curve   is fitted by the power-law increase $P(t) = 1-C(t-a)^{-\nu}$,
where $C\approx 0.8$, $a$ is the respective shift (see Fig.\ 7) and the
exponent $\nu \approx$ 2, 1.75, and 1.68, for $f=$ 2.4, 3.4, and 4.4,
respectively. Again, the characteristic exponent $\nu$ decreases when the
pinning becomes stronger, thus contributing to the slow dynamics but
within a power-law, rather than a stretch-exponential dependence.

\section{Conclusions and discussion}

We have shown that, in addition to the  measurements of the hysteresis
loop properties and smooth component of the switching current, the domain
structure in a ferroelectric
sample can be indirectly determined by the non-invasive dynamic
measurements of the switching current noise and back-switching processes.
Fractal character of the dynamic response indicates that a nontrivial
domain structure can occur under slow field driving.
The measured properties in these processes can be expressed
in terms of universal  exponents, which turn to be related in a unique way
to the underlying structure and motion of domain walls.
We have shown how the universal features of the switching current noise
are quantified through two distinct sets of scaling exponents.  The
universal scaling behavior is associated  with a dominant scenario of the
domain wall dynamics: either the motion and depinning of large  domain
walls, in the case of weak pinning, or  bulk nucleation and growth of many
small 3-dimensional domains, in the case of strong pinning.
The  pinning in our model is provided by the local electric random fields.

Similarly, the polarization reversal in a strong applied field,
and back-switching of the domains when the field is set to zero, are quick
when large domains occurs, whereas they are slow when  many small domains
are nucleated. In the latter case the polarization relaxes according to
a stretch-exponential law (the stretching exponent $\sigma <1$) whereas it
reverses with a fractal power of time (the characteristic exponent $\nu <2$)
 when a large opposite field is switched on.
In the case of weak pinning, on the contrary, the relaxation is exponential
($\sigma \approx 1$) and the reversal  algebraic in time with $\nu =2$.
Dependence of the stretching exponent $\sigma (f)$ on the pinning strength
is in qualitative agreement with measurements of back-switching
in doped ${\mathrm{SBN:Ce^{3+}}}$ relaxor ferroelectrics \cite{Kleemann},
again suggesting that the  local electric fields  contribute
significantly to slow the dynamics in this class of relaxors \cite{Kleemann}.
(Note that the characteristic time appears unrealistically small in our
lattice model.) We are not aware of measurements of temporal  behavior of
the polarization reversal and of the switching current noise in these systems.

The observed  scaling behavior of the switching current depends on
the  rate at which the driving field changes. We have demonstrated that
in both universality classes the scaling exponents decrease {\it
logarithmically} with increased driving rate $\Delta E/J$. This is in
sharp contrast to the cases where the dominant type of disorder is due to
random vacancies in spin-systems \cite{BT_summary} and other avalanche-type
dynamic systems driven at a finite rate \cite{TP}, where the scaling
exponents decrease and fractal dimensions increase {\it linearly} with the
driving rate.
Therefore, occurrence of the logarithmic dependences  of the scaling
exponents on the driving frequency may serve as an additional test of
the presence of the real random fields in  ferroelectrics.

Throughout this work we used the simulation algorithm at finite driving
 rates $\Delta E >0$, that mimics the  realistic situation in experiments.
The obtained set of scaling exponents, together with their
dependences on the driving rate, belongs to one (of the two) universality
classes, which implies  a definite domain structure of the system, and hence
the expected  character of the switching processes. Our results suggest that
a crossover from one to the other universal behavior both in the switching
current noise and in the polarization reversal and relaxation can be achieved
by varying  the strength of pinning due to random electric fields.
As mentioned above, in practice there are various ways
to accomplished the crossover by reducing the pinning in ferroelectrics
\cite{Shur_et_al,Kleemann,PLZT-HL}.
Here we have not discussed the properties of the dynamic phase transition
separating these two scaling behaviors at a critical value $f_c$, nor the
critical exponents describing the depinning transition along the critical
field $H_d(f)$ for $f<f_c$.
This questions require an additional study and proper finite size scaling
analysis, for instance along the lines as done for the two-dimensional
case in Ref.\cite{TN}.
Finally, by allowing a frequent appearance of the unscreened depolarization
 fields  a  finite probability  of negative pulses in the switching current
builds up ($b\neq 0$). The presence of negative pulses slows the switching
process, interferes with the effects of finite driving rates, and alters
the scaling properties of noise. A more complete analysis of the switching
processes for non-vanishing probability $b\neq 0$ is left for a future
study.

Finally, it should be stressed that, in spite of its simplicity, the proposed
model seems reasonable for dynamic properties of three-dimensional uniaxial
relaxor ferroelectrics.
Whereas, in the single-crystalline normal ferroelectrics a  more realistic
modeling of the effects of interactions and depolarization fields
on domain structure is required. Nevertheless, we believe that, when
restricted only to  the {\it universal} scaling behavior of the noisy
component of the switching current, the presented results for the case of
large domains may not depend significantly on the details of interactions.
Further research on the subject is necessary.

\acknowledgments
Work is supported by the Ministry
of Education, Science and Sports of the Republic of Slovenia.

 \narrowtext
\begin{table}
  \begin{center}
    \begin{tabular}{|c|c|c|c|}
      $f$ (pinning) &Exponent& 3D & 2D\\
\hline \hline$f\approx f_c$& $\tau _s$& 1.78 & 1.54\\
(weak)& $\tau _t$& 2.20 & 1.83\\
\hline$f> f_c$& $\tau _s$& 1.66 & 1.30\\
(strong)& $\tau _t$& 2.03 & 1.47\\
    \end{tabular}
  \caption{Scaling exponents for weak pinning ($f\approx f_c$) and for strong
pinning ($f> f_c$) for 3-dimensional (3D) system at driving rate $4\times
10^{-4}$,  results of this work, and for
2-dimensional (2D) system  at zero driving rate,  from Ref.\ [25].
    Estimated error bars  are within $\pm 0.03$.}
  \end{center}
  \label{table1}

\narrowtext
\begin{figure}[b!] 
\epsfxsize=80mm\epsffile[46 72 513 488]{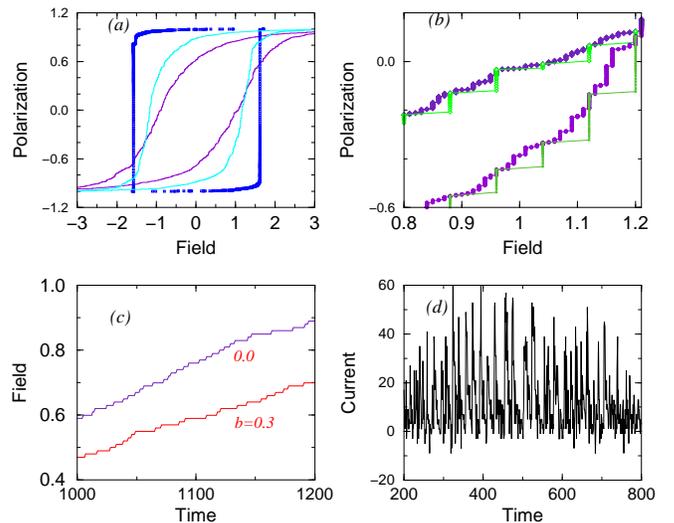}
\caption{(a) Hysteresis loop for driving rate $r\equiv \Delta E/J=$0.01 and
varying strength of disorder $f=$4, 3.2, and 2 (inside out).
(b) Sections of the first two loops from (a) for two different
driving  rates $r=$0.01 (dark) and $r=$0.08 (clear symbols).
(c) Variation of the driving field with time and (d) the electrical Barkhausen
noise (switching current) for finite probability of back flips $b=0.3$ .
Fig.\ 1 from Ref.\ [18].}
\label{fig1}
\end{figure}

\begin{figure}
\epsfxsize=80mm\epsffile[58 63 509 639]{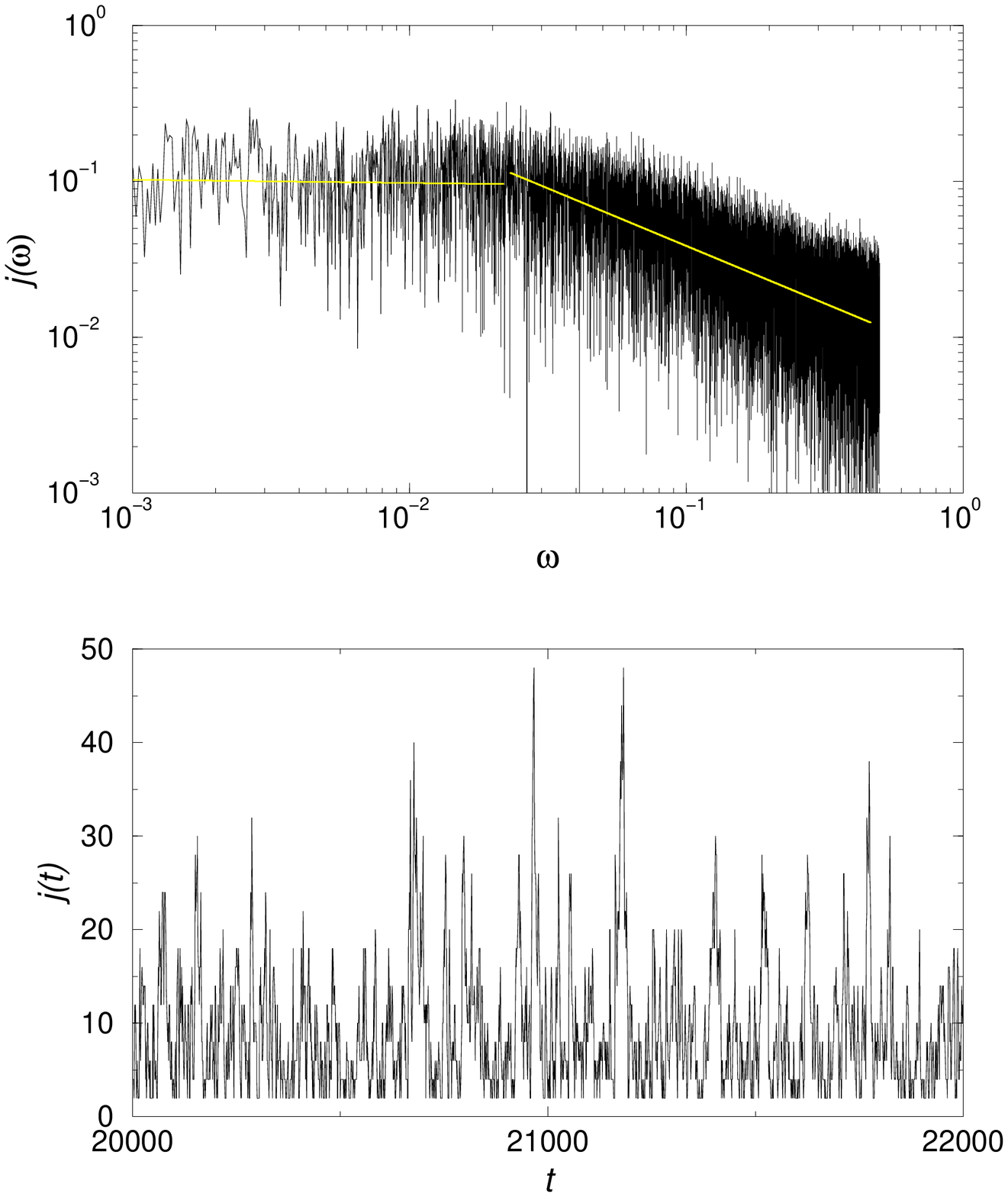}
\caption{Part of the switching current signal $j(t)$ against cumulative
time $t$ (lower panel) and the Fourier spectrum of the signal (top panel)
for moderate pinning $f=3.6$ . Slope of the fitted area
$\phi = 0.738\pm 0.007$ . Only nonzero values of current are shown (see text).}
\label{fig2}
\end{figure}

\begin{figure}
\epsfxsize=80mm\epsffile[45 72 507 670]{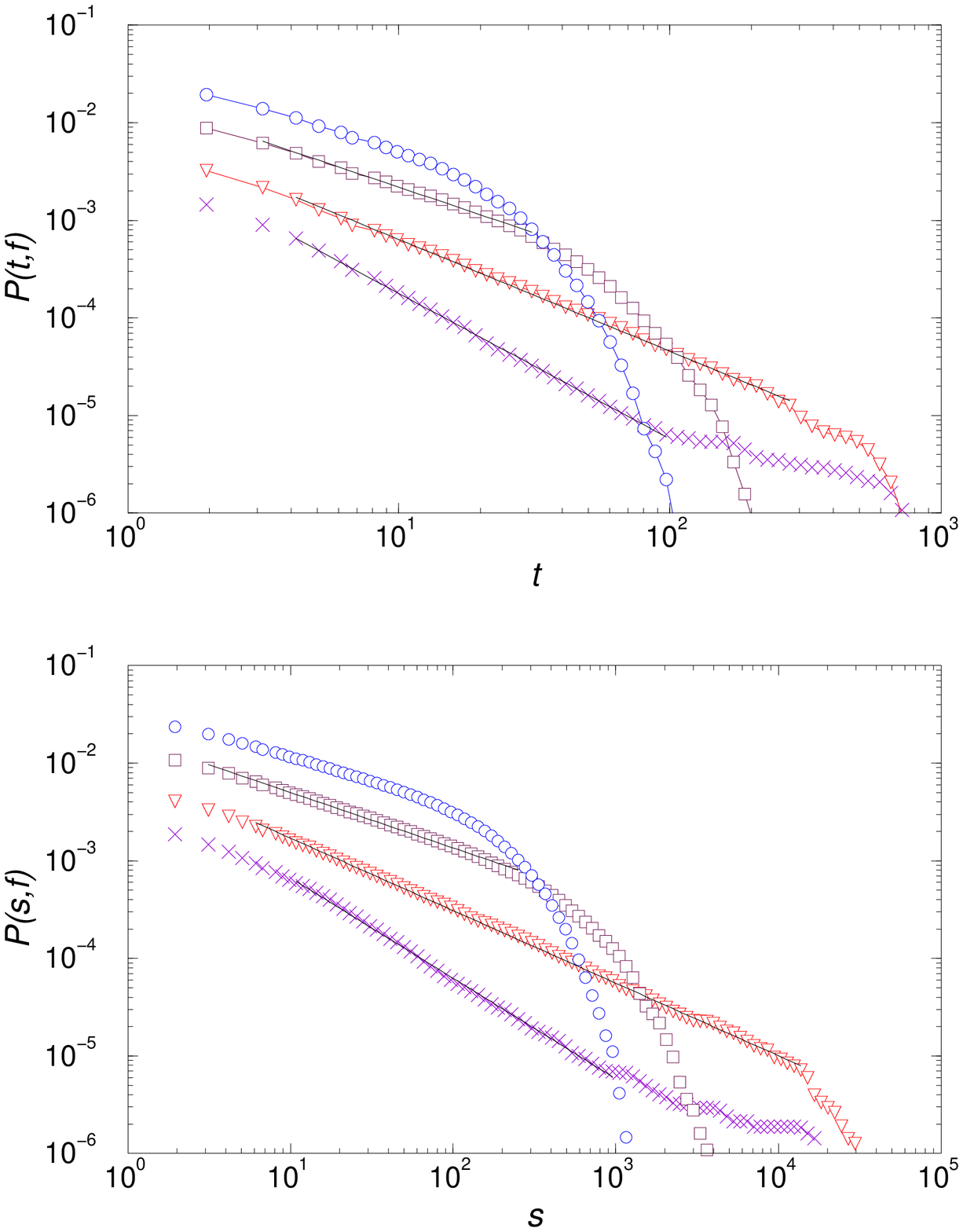}
\caption{ Cumulative distributions of switched domain size $P(s,f)$ vs. size
 $s$ (lower panel) and duration of switching $P(t,f)$ vs. duration $t$
(top panel) for
varying strength of pinning $f$ =2.2, 2.4, 2.8 and 3.2 (bottom to top, both
panels). Driving rate is fixed $\Delta E/J$ = 0.001. Fitted values of three
slopes, corresponding to full lines, are (bottom to top): $\tau_s-1$=1.029(4),
0.747(2), 0.568(5), on the lower panel, and $\tau_t-1$= 1.482(6), 1.145(7),
and 0.94(2), on top panel.
In total $4.32\times 10^6$ spins divided into 20 different samples was
used in simulations. Distributions are normalized by the actual
number of avalanches during the full cycle and the number of samples.
Data are log-binned with ratio 1.1  .}\label{fig3}
\end{figure}

\begin{figure}
\epsfxsize=82mm\epsffile[44 72 507 670]{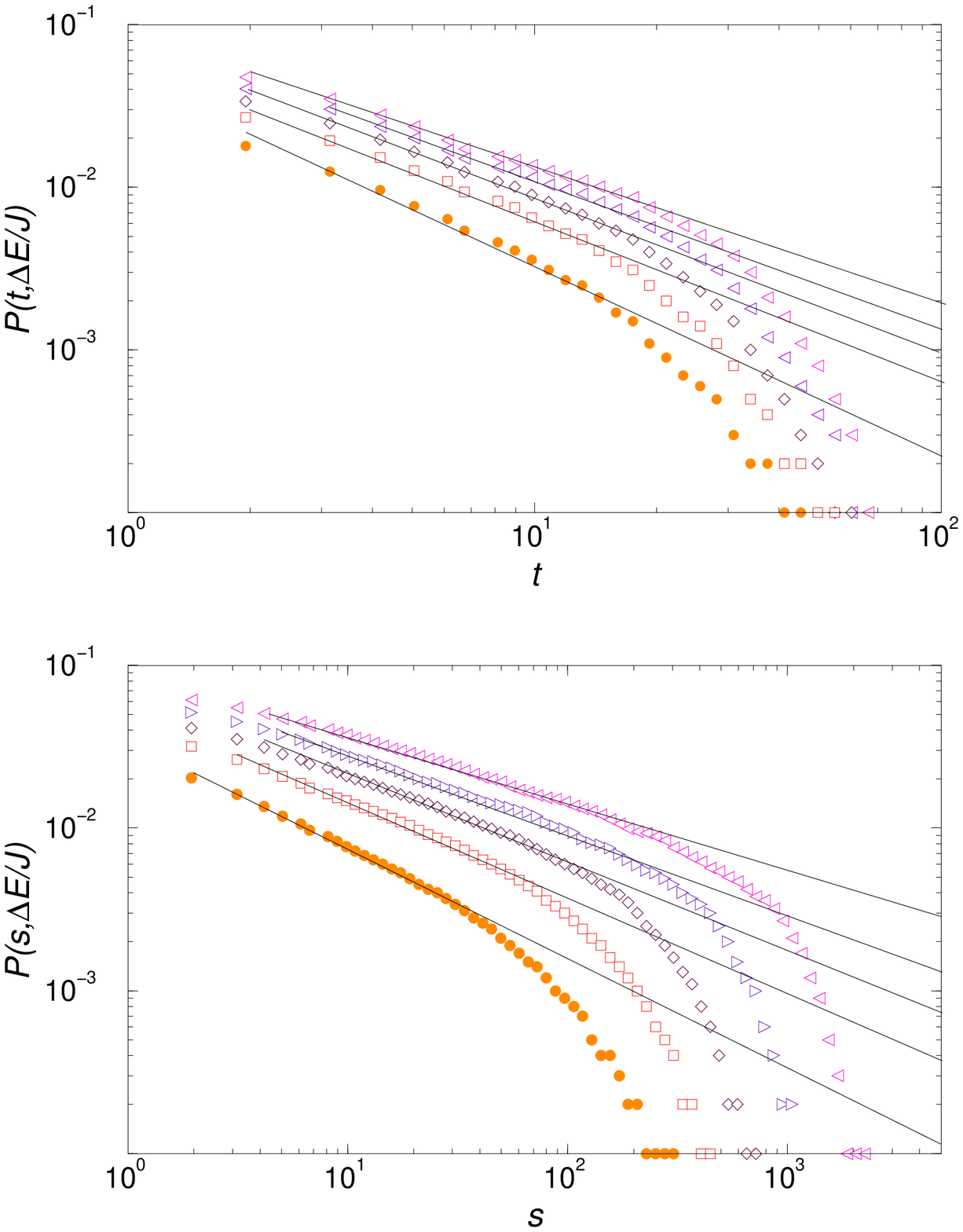}
\caption{Cumulative distributions of switched domain size $P(s,\Delta E/J)$
vs. $s$ (lower panel) and duration of switching $P(t,\Delta E/J)$ vs. $t$
(top panel) for fixed pinning $f=3.2$ and varying driving rate $\Delta E/J$
= $4\times10^{-4}$,
$1\times10^{-3}$, $2\times10^{-3}$, $4\times10^{-3}$ and  $8\times10^{-3}$
(bottom to top on both panels). Normalization  to the actual
number of avalanches results in the vertical shift of the curves.
Scaling region indicated by the section of the straight line.
 Log-binning ratio 1.1 .}\label{fig4}
\end{figure}

\begin{figure}
\epsfxsize=82mm\epsffile[41 72 507 670]{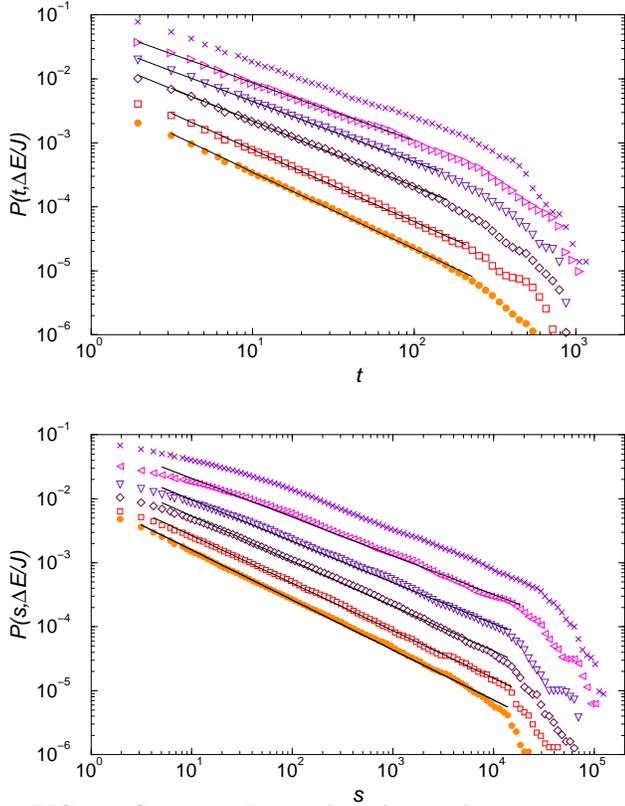}
\caption{Same as Fig.\ 4 but for weak pinning regime $f=2.4$. Driving rates
are (bottom to top)  $\Delta E/J$ = $4\times10^{-4}$,
$1\times10^{-3}$, $2\times10^{-3}$, $4\times10^{-3}$, $8\times10^{-3}$
and $1.2\times10^{-2}$. Fitted area on each curve is shown by solid line. }
\label{fig5}
\end{figure}

\begin{figure}
\epsfxsize=82mm\epsffile[70 93 324 385]{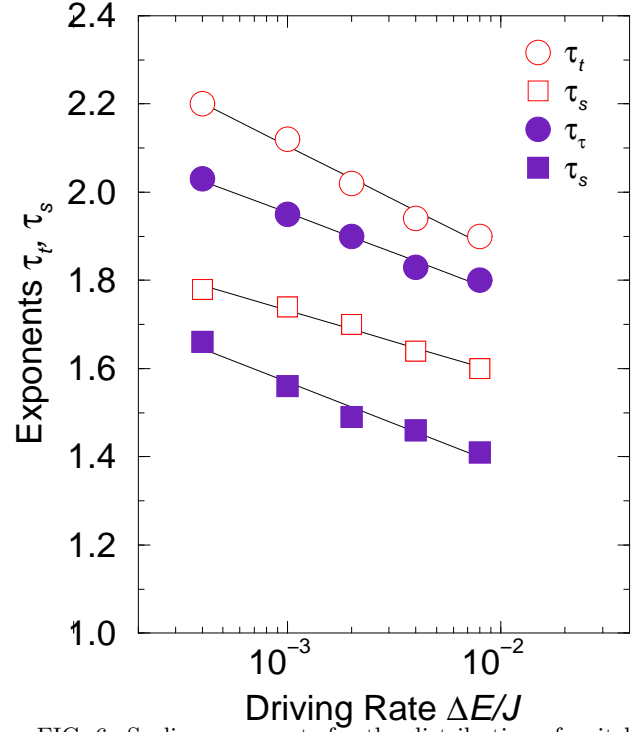}
\caption{Scaling exponents for the distribution of switched area (squares)
and duration of pulses (circles) against driving rate $\Delta E/J$.
Filled and open symbols correspond to strong  and weak pinning, respectively.
 Fit lines:  $\tau _X= A_X -B_X\ln (\Delta E/J)$, see text.}
\label{fig6}
\end{figure}

\begin{figure}
\epsfxsize=80mm\epsffile[53 72 508 412]{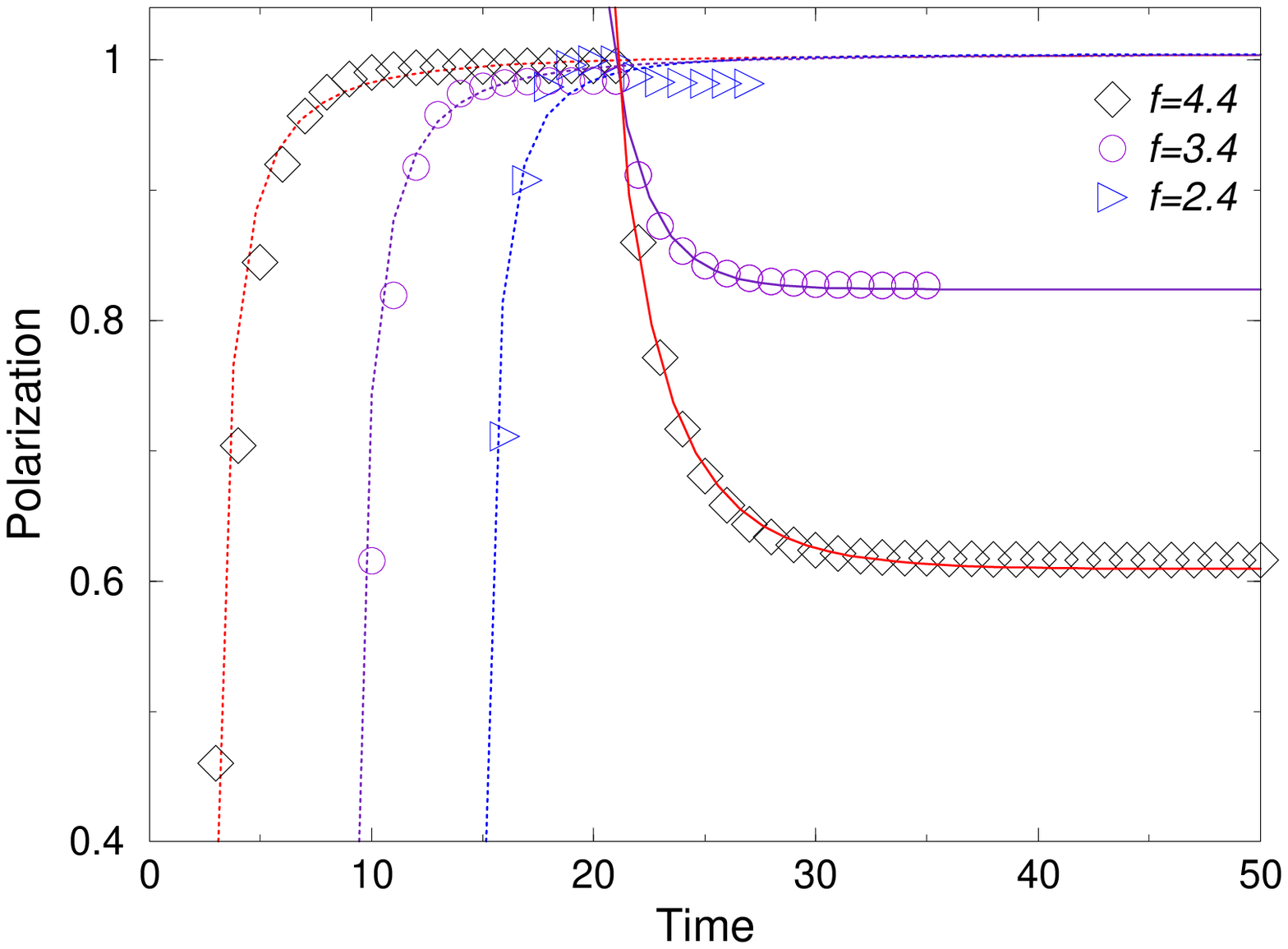}
\caption{Polarization reversal at field $E/J=3$ (left part) and relaxation
after field was set to zero (right part) against time for three values of
pinning strength $f$, as indicated. Latter two lines were shifted to the right
for better display.  Fits according to $P=1-C(t-a)^{-\nu}$ (dotted lines),
and $P=A+B\exp[-(t/2-11)^\sigma]$ (full lines), see text. }
\label{fig7}
\end{figure}

\end{table}
\end{multicols}

\begin{references}
\bibitem{comment_planarDW}Apart from weak pinning, more realistic
interactions, e.g.,  including elastic
coupling are responsible for occurrence of the plane domain walls in
 normal ferroelectrics such as  ${\mathrm{BaTiO_3}}$.
\bibitem{comment_integration}
Notice that the properties of noise vary with the applied field, see for
instance Refs.\ \cite{Shur,FTT}.
\end{references}
\end{document}